# Confirmation of room-temperature long range magnetic order in GaN:Mn

Y. H. Zhang[1], Z. Y. Lin[1], F. F. Zhang[1], X. L. Yang[1], D. Li[1], Z. T. Chen[1], G. J. Lian[1], Y. Z. Qian[1], , X. Z. Jiang[1], T. Dai[1], Z. C. Wen[2], B. S. Han[2], C. D. Wang[1], G. Y. Zhang[1, a)]

[1] *State Key Laboratory of Artificial Microstructure and Mesoscopic Physics, School of Physics, Peking University, Beijing 100871, People's Republic of China*

[2]*State Key Laboratory of Magnetism, Beijing National Laboratory for Condensed Matter Physics, Institute of Physics, Chinese Academy of Science, Beijing 100190, People's Republic of China*

## *Abstract*

We propose a method for nano-scale characterization of long range magnetic order in diluted magnetic systems to clarify the origins of the room temperature ferromagnetism. The GaN:Mn thin films are grown by metal-organic chemical vapor deposition with the concentration of Ga-substitutional Mn up to 3.8%. Atomic force microscope (AFM) and magnetic force microscope (MFM) characterizations are performed on etched artificial microstructures and natural dislocation pits. Numerical simulations and theoretical analysis on the AFM and MFM data have confirmed the formation of long range magnetic order and ruled out the possibility that nano-clusters contributed to the ferromagnetism. We suggest that delocalized electrons might play a role in the establishment of this long range magnetic order.

[a)]Author to whom correspondence should be addressed. Electronic mail: gyzhang@pku.edu.cn

Over the last several years, diluted magnetic semiconductors (DMS) have attracted increasing attention due to its potential application in spintronics devices. Stimulated by predictions of robust ferromagnetism in a variety of systems [1], the observation of a ferromagnetism well above room temperature (RT) has been reported for several diluted magnetic oxides (DMO) and III-V DMSs [2], including the Mn-doped GaN with the highest Curie temperature up to 940K [3-5]. However, as for GaN:Mn, there are many significantly discrepant experimental results as anti-ferromagnetic [6], spin-glass [7] behavior and also a paramagnetism in the most recently reported single-phase $Ga_{1-x}Mn_xN$ grown by metal organic vapor phase epitaxy (MOVPE) [8].

For the RT ferromagnetic DMSs and DMOs, the origin of the ferromagnetism has begun to emerge as one of the most unsettled problems, which is attributable to the lack of a magnetic characterizing method at the nanoscale. Most of the reports that discussed observation of ferromagnetism primarily based on magnetic hysteresis measurement, which could not clarify magnetic origins for weak magnets in DMS systems. It has been found that there are many mechanisms that could be responsible for the signals of magnetic hysteresis, such as doped magnetic matrices[9], ferromagnetic clusters [10], second phases [11,12], substrates [13], interface [14], and other external pollution [15]. A more effective magnetic characterizing method for DMS and DMO could be achieved by using magnetic force microscope (MFM) for studying nanoscale ferromagnetic structures. However, besides the commonly reported MFM study on local magnetic microstructures [16], there has been no convincing MFM study on the long range magnetic order in RT DMS reported so far, due to the undetectably weak surface stray magnetic field in DMS thin films.

In this article, a method was proposed for the nanoscale magnetic characterization of GaN:Mn thin films grown by metal organic chemical vapor deposition (MOCVD), using MFM and atomic force microscope (AFM) measurements on etched artificial structures and natural dislocation pits. Long range magnetic order without Mn rich nano-clusters was then confirmed through the simulations and analysis on MFM and AFM results.

The GaN:Mn films were epitaxially grown by MOCVD on GaN substrates, using Bis (methylcyclopentadienyl) manganese ($MCP_2Mn$) as manganese source, and $H_2$ as carrier gas, respectively [17]. The films were n-type as determined by Hall analysis. The chemical composition of the thin film was examined by proton induced X-ray emission (PIXE) and secondary ion mass spectroscope (SIMS) technique. Fig. 1(a) shows the PIXE spectrum of the GaN:Mn sample, which reveals the Mn concentration to be 3.8% [17].The inset of Fig. 1(a) is the result of SIMS measurements, indicating the uniform distribution of Mn composition along the growth direction.

The chemical and structural properties of the GaN:Mn thin films were studied by using X-ray diffraction (XRD), transmission electron microscope (TEM), x-ray absorption near-edge structure (XANES) and extended x-ray absorption fine structure (EXAFS). Presented in our previous work, detailed XRD [5] and TEM [18] both reveal that no secondary phases were found within the experimental detection limits. The XAFS experiments were performed at beamline 1W1B at the Beijing Synchrotron Radiation Facility (BSRF) in Beijing, China. The XANES spectra were recorded at the Mn K edge (6539 eV) in the fluorescence mode. The K edge x-ray absorption spectrum of Mn in GaN:Mn is shown as the experimental curve in Fig. 1 (b). Theoretical XANES spectra were simulated using the multiple-scattering formalism within the muffin tin approximation [19], where the best fits were obtained by employing a substitutional model of one Mn atom at a Ga site, as the theoretical curve for $Mn_{Ga}$ incorporation shown in Fig. 1 (b). As a sensitive tool for the site identification and valence state of Mn, the EXAFS spectrum is shown in the inset figure of Fig. 1(b). The notable Mn-Ga peak supports the $Mn_{Ga}$ incorporation and more simulations to confirm this incorporation in the EXAFS will be presented elsewhere. The XRD, TEM, XANES and EXAFS results all indicate the absence of second phases and the substitutional Mn occupation in Ga sites in the GaN:Mn thin films. To explore the possibilities of this single-phase GaN:Mn with the Mn concentration up to 3.8%, first-principle calculations were carried out based on the theory developed by Zhang *et al* [20-22]. Taking into account the role of growth surface [22] and the H passivation [23], the substitutional Mn of several atom percent higher than 4% can be achieved.

The magnetization measurements were performed by a superconductor quantum interference device (SQUID). The magnetic hysteresis measured at 300K [Fig. 1(c)] and at 380K [the inset (a) of Fig. 1(c)] demonstrate a ferromagnetism with the Curie temperature higher than 380K (the upper temperature limit of the SQUID is 400K). The inset (b) of Fig. 1(c) shows the $H_c$ at different temperature indicating virtually no remnant magnetization.

Our MFM measurements were performed by a Digital Instruments NanoScope Ⅲa-D3000

magnetic force microscope with the spatial resolution of 20 nm. Typical MFM image recorded on the surface of as-grown GaN:Mn sample is shown in Fig. 2(a). The MFM images all reveal random weak response independent of any grain-like clusters when scanning in the whole surface. To overcome the difficulty of detecting weak stray magnetic field, the isolated GaN:Mn islands [Fig. 2(c)] with trapezium-shaped cross section were fabricated by using focused ion beam (FIB) to etch down to the GaN layer below. The FIB etching experiments were carried out in a FEI DB235 system with the accelerated voltage and the ion-beam current of Ga ion beam fixed at 30kV and 100 pA respectively. Since the Gaussian energy distribution of the beam caused the edge of the etched structures to be worn [24], the surface of the structures was actually close to sine-shape and smooth, as shown in the AFM image [Fig. 2(c)]. Elaborate PIXE, SIMS and SQUID investigations reveal the consistency of sample chemical properties and ferromagnetism before and after patterning via FIB.

The corresponding MFM image of the structure is shown in Fig. 2(d). Noting the difference from the topography image and lacking steep steps, the magnetic signals are real and not supportive of the presence of any magnetic clusters [9]. High contrast magnetic responses exhibit an anti-symmetric dark-bright pattern in favor of long range magnetic order, which is quantitatively confirmed by the following computation.

The computation is performed on the MFM response, which measured the total force gradient acted on the magnetic tip [25]. The idea of magnetic charge density is employed, which is a useful mathematical construct in the analysis and calculation. The surface magnetic charge density could be written as $\sigma_\mathrm{m} = \mu_0 M \cos\theta$, where $M$ is the magnetization intensity, and $\theta$ is the angle between the direction of $M$ and a unit vector normal to the surface.

Due to the quite small distance between the tip and the surface as below 25 nm experimentally required for the detection of weak DMS magnetic response, the MFM response is reasonably assumed as directly proportional to the magnetic charge density of the surface spot below the tip. In this idea, the significance of the artificial structure lies in producing the variation of $\theta$ and therefore the stronger contrast of the MFM response.

The numerical simulation was performed under the assumption of uniform magnetization using a 1024×1024 three-dimensional data packet extracted from the AFM result and the corresponding 1024×1024 MFM data packet. The MFM data referred to the distribution of magnetic charge density [Fig. 3(a)] as discussed before. The normal vector of each AFM unit was derived by simulating a plane at each unit through the method of least squares. Then the angle $\theta$ was obtained and the distribution of magnetic charge density was calculated shown in Fig. 3(b). As can be seen, the computational and experimental distribution is in a good agreement. They have nearly the same zero points and the distribution of peaks and valleys. Thus, the assumption of the computation, the long range magnetic order and uniform magnetization in the GaN:Mn, has been substantiated.

Due to certain similarity between the artificial microstructures and the natural dislocation in their great roughness, the MFM investigation on natural dislocation pits was expected to show observable signals. An area is found with well dispersed hexagonal dislocation pits similar to the regularly reported threading dislocation in epitaxially grown GaN [26,27] and GaAlN [28]. The AFM and MFM images of this area are shown in Fig. 4(a) and Fig. 4(b). The AFM image focused on several single pits [Fig. 4(c)] reveals the typical dislocation shape [Fig. 4(d)] through the line profile analysis. Its corresponding MFM image [Fig. 4(e)] reveals a double-polar anti-symmetric

MFM pattern (a dark "south" edge and a yellowish "north" edge) for an isolated pit (It is also notable that no magnetic responses are observed at the flat bottom area of dislocation pits). The exhibition of the same polarity direction of these MFM patterns shown in Fig. 4(b) demonstrates the uniform magnetization and long range magnetic order in the GaN:Mn. This is confirmed by computations on the assumption of uniform magnetization. Two typical simulated MFM patterns are shown in Fig. 4(f) and Fig. 4(g) with the different directions of the magnetization, where the pattern with magnetization direction as shown in the latter is basically consistent with the MFM pattern of isolated dislocation pits shown in Fig. 4(e).

As no grain-like patterns found in the searches either covering the whole MFM data of etched microstructure shown in Fig. 3(a) where the resolution is below 30 nm, or at the bottom of dislocation pits where the resolution is below 20 nm, the nano-clusters with a size above MFM's spatial resolution could be ruled out. If there exists Mn rich nano-clusters below the MFM's spatial resolution, the sample would behave as a superparamagnetism, and the size of hypothetic magnetic clusters can be estimated by $d \approx (150 k_B T_B / \pi K)^{\frac{1}{3}}$ [29, 30], where $k_B$ is the Boltzmann constant, $K = M_S H_A / 2$ is density of anisotropy energy, and $T_B$ is the blocking temperature. The monotonously decreasing $H_c$ from 110 Oe at 5 K to 55 Oe at 380 K indicates the hypothetic magnetic clusters should be single domain [31]. Then the anisotropy field, $H_A$, can be identified with the low-temperature $H_c$, about 100 Oe, leading to $d \approx 122$ nm for $T_B = 380$K. According to Fig. 1(c), $T_B$ should be higher than 380K and consequently $d$ should be larger than 122 nm. Such a size is far above the MFM's spatial resolution, in contradiction to the hypothesis. Therefore, the Mn rich nano-clusters contributing to ferromagnetism measurements could be ruled out in our GaN:Mn samples.

Most recently, *Dietl* group [8] reported the paramagnetic behavior of the single-phase epitaxially grown $Ga_{1-x}Mn_xN$ with x up to 1%. In contrast, our epitaxially grown $Ga_{1-x}Mn_xN$ with x up to 3.8% has been demonstrated as single-phase, cluster-free as well as magnetic-ordered. Among various controversial models [32], our observed RT ferromagnetism is difficult to be explained by the free hole-mediated magnetism based on Zener's *p-d* exchange, due to the absence of sufficiently high concentrations of both substitutional magnet impurities (above 5%) and valence-band holes (above $3.5\times10^{20}cm^{-3}$). Instead, our previous reports [33-36] on electrical, structural and optical properties of GaN:Mn revealed a consistence in details with the double exchange mechanism developed by Sato, Zunger *et al* [37-40]. The only pity lies in that Sato *et al*'s model seems difficult to explain the high $T_c$ in the GaN:Mn with such a content of Mn not so high as expected. In this GaN:Mn, how a short-ranged double exchange mechanism contributes to a long-range magnetic order is not clear. Our previous reported [35, 36, 41] and some recent experiments revealed that the delocalized electrons related to certain point defects induced by the Mn-doping might play a significant role in the establishment of long-range magnetic order. Further experimental and theoretical works in this field are needed and we will also specifically discuss these in our future papers.

In summary, a method is forwarded for nano-scale characterization of long range magnetic order in diluted magnetic systems. Experiments including PIXE, SIMS, EXAFS, TEM and XANES reveal the single-phase MOCVD grown GaN:Mn films with the concentration of Ga-substitutional Mn up to 3.8%. We demonstrate the formation of long range magnetic order independent of nano-clusters in the GaN:Mn using AFM and MFM measurement with related simulations and analysis. These results exclude a variety of controversial origins of room temperature ferromagnetism in our GaN:Mn thin films, including clusters, second phases, substrates, interface and other surface contaminations. We suggest that delocalized electrons might contributed to the establishment of this long range magnetic order

**ACKNOWLEDGMENTS**

We thank Prof. Fernando Ponce, Prof. Yunfeng Xiao and Doc. Lei Wang for helpful discussions. This work was supported by the National Natural Science Foundation of China (Grant Nos. 60577030, 60776041, and 60876035), National Key Basic Research Special Foundation of China (Grant Nos. TG2007CB307004 and 2006CB921607).

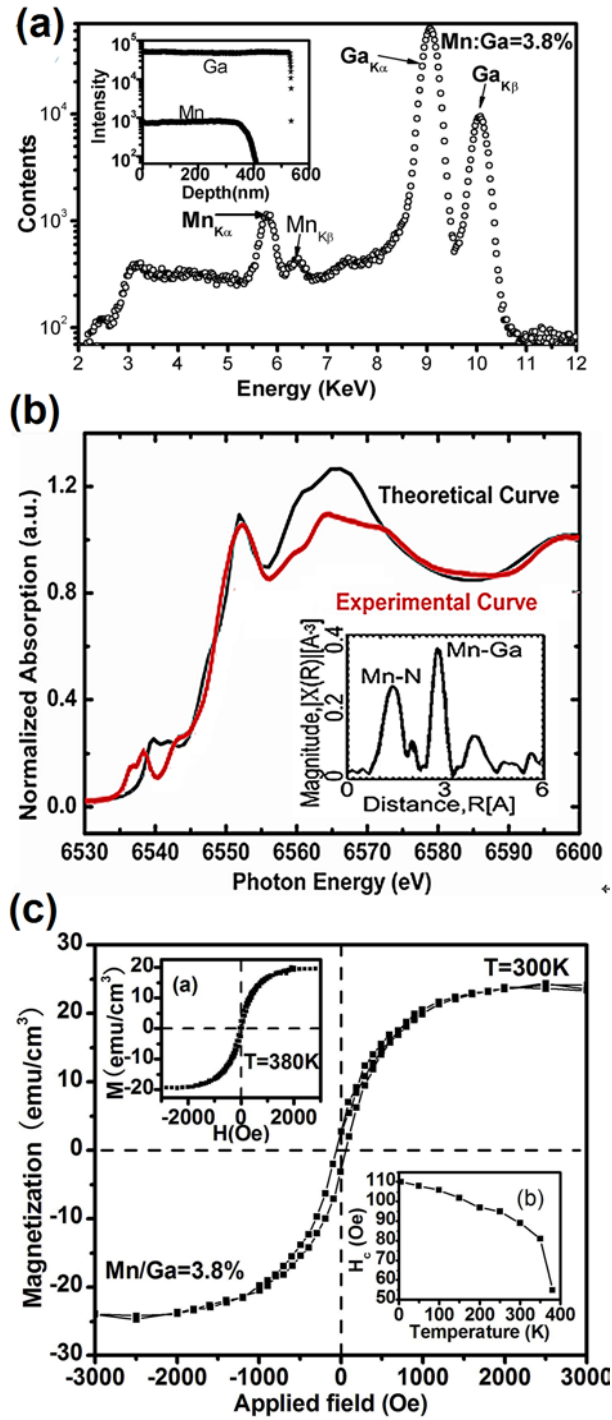

FIG. 1. (color online) (a) the PIXE spectrum of the GaN:Mn sample. The inset figure is the result of SIMS measurements. (b) the experimental curve as the K edge x-ray absorption spectrum of Mn in GaN:Mn and the theoretical curve as the fitting employing a substitutional model of one Mn atom at a Ga site. The inset figure is the EXAFS spectrum of the GaN:Mn. (c) Magnetization of GaN:Mn films measured by SQUID. The inset (a) is the magnetic hysteresis at 380 K with coercivity force ($H_c$) of 55 Oe. The inset (b) is the $H_c$ at different temperature.

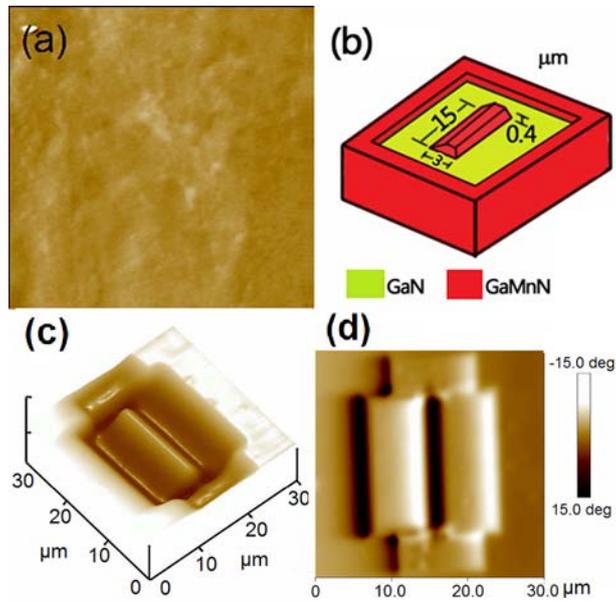

FIG. 2. (color online) (a) MFM image (phase mode, Z 10 degree) recorded on a 5×5 μm$^2$ area for the as-grown GaN:Mn film fabricated by MOCVD. The schematic figure (b) and corresponding AFM image (c) of the artificial structure, which is etched by FIB with the ion-beam current fixed at 30kV and 100 pA. (d) the MFM image of the structure (phase mode, Z 30 degree) recorded on the 30×30 μm$^2$ area.

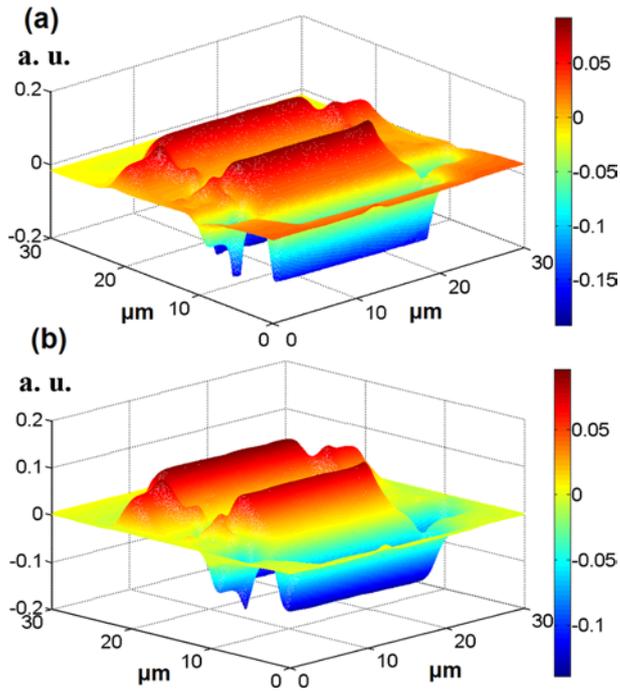

FIG. 3. (color online) (a) the distribution of magnetic charge density covering the whole structure obtained from the 1024×1024 MFM data packet. Computational distribution (b) agrees well with the experimental result (a) from MFM.

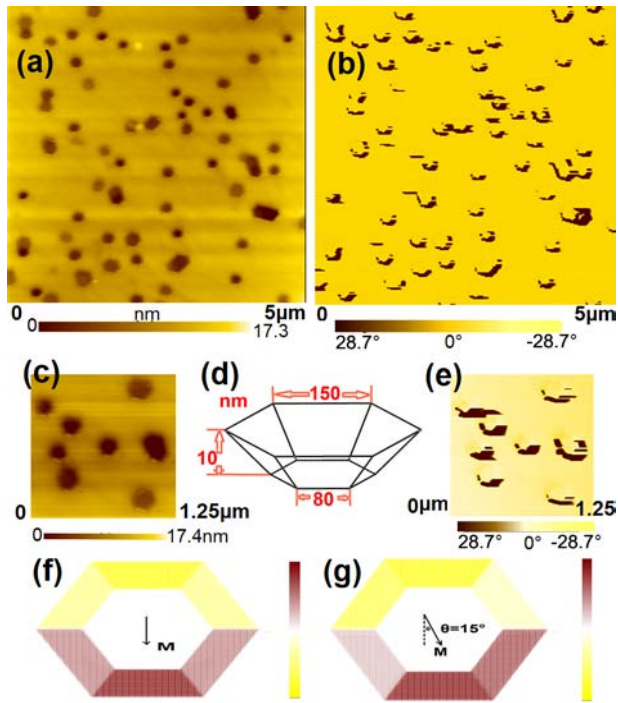

FIG. 4. (color online) the AFM image (a) and the MFM image (b) of the dislocation pits recorded on a 5×5

$\mu m^2$ area. The AFM image (c) and the MFM image (e) of the dislocation pits recorded on 1.25×1.25 $\mu m^2$ area. The schematic illustration of the typical shape of dislocation pits (d) used in the computation. The computational MFM pattern of the dislocation pits (f) and (g) with different angles between the magnetization and the dislocation pits.